\documentclass{elsart}
\usepackage[dvips]{epsfig}
\def \be{\begin{equation}}
\def \ee{\end{equation}}
\def \bdm{\begin{eqnarray}}
\def \edm{\end{eqnarray}}

\begin{document}
\begin{frontmatter}
\title{Finite Gyroradius Corrections in the Theory of Perpendicular Diffusion\\1. Suppressed Velocity Diffusion}
\author{A. Shalchi}
\address{Department of Physics and Astronomy, University of Manitoba, Winnipeg, Manitoba R3T 2N2, Canada, tel: 1-204-474-9874, fax: 1-204-474-7622}
\ead{andreasm4@yahoo.com}
\begin{abstract}
A fundamental problem in plasma physics, space science, and astrophysics is the transport of energetic particles interacting
with stochastic magnetic fields. In particular the motion of particles across a large scale magnetic field is difficult to
describe analytically. However, progress has been achieved in the recent years due to the development of the unified non-linear
transport theory which can be used to describe magnetic field line diffusion as well as perpendicular diffusion of energetic particles.
The latter theory agrees very well with different independently performed test-particle simulations. However, the theory is
still based on different approximations and assumptions. In the current article we extend the theory by taking into account
the finite gyroradius of the particle motion and calculate corrections in different asymptotic limits. We consider different
turbulence models as examples such as the slab model, noisy slab turbulence, and the two-dimensional model. Whereas there
are no finite gyroradius corrections for slab turbulence, the perpendicular diffusion coefficient is reduced in the other two cases.
The matter investigated in this article is also related to the parameter $"a^2"$ occurring in non-linear diffusion theories.
\end{abstract}
\end{frontmatter}
\section{Introduction}
Energetic particles interact with turbulent magnetic fields while they propagate through a magnetized plasma. Due to this
interaction, their motion is a stochastic motion. In addition to turbulent fields $\delta \vec{B}$, there are also large
scale magnetic fields $\vec{B}_0$ influencing the particle orbit. The latter field is usually called guide field, mean field,
background field, or large scale field. This type of configuration can be found in different physical environments
such as fusion devices, the solar wind, or the interstellar medium (see, e.g., Schlickeiser 2002, Spatschek 2008, and
Shalchi 2009 for reviews). The turbulent fields described above lead to a diffusive motion of the particles. Due to
the large scale field one has to distinguish between diffusion along and across that field. It is often assumed that
perpendicular diffusion is more difficult to describe analytically compared to parallel transport. It should be noted
that sub- and superdiffusive transport has been discussed more recently in the literature (see, e.g., Zimbardo et al. 2006,
Pommois et al. 2007, Shalchi \& Kourakis 2007, and Zimbardo et al. 2012).

Previous theories for perpendicular diffusion such as the Non-Linear Guiding Center (NLGC) theory of Matthaeus et al. (2003)
and the Unified Non-Linear Transport (UNLT) theory of Shalchi (2010) neglect the rotation of the particle in the direction
perpendicular with respect to the large scale field. Furthermore, as equation of motion in such theories, the following {\it Ansatz}
is used
\be
V_x = a v_z \delta B_{x} \left[ \vec{x} (t), t \right] / B_0.
\label{introa}
\ee
Here we have used the particle position $\vec{x} (t)$ at time $t$, the $z$-component of the particle velocity vector $v_z$,
the $x$-component of the turbulent magnetic field $\delta B_x$, the mean magnetic field $B_0$, and the $x$-component of the
guiding center velocity vector $V_x$ (see Section 2 for more details). The parameter "a" used here can be seen as an unknown
parameter. If non-linear theories are compared with test-particle simulations, best agreement is usually found for values
between $a^2 = 1/3$ and $a^2 = 1$ (see, e.g., Matthaeus et al. 2003 and Tautz \& Shalchi 2011). However, analytical work
based on the Newton-Lorentz equation suggests that this parameter is between $a^2 =1$ and $a^2 =2$ (see Shalchi \& Dosch 2008,
Dosch \& Shalchi 2009, and Dosch et al. 2013).

Especially the UNLT theory has shown remarkable agreement with test-particle simulations for different turbulence models
such as the slab/2D model, the Goldreich-Sridhar model, Alfv\'en waves, and noisy reduced MHD turbulence (see Tautz \& Shalchi 2011,
Shalchi 2013, Hussein \& Shalchi 2014, Shalchi \& Hussein 2014, Shalchi \& Hussein 2015). Therefore, we conclude that the main
problem in the theory of perpendicular transport is Eq. (\ref{introa}) and therewith the parameter "a".

In the current paper we explore the influence of finite gyroradius effects. To do this we employ two different approaches, namely
\begin{enumerate}
\item Quasi-Linear Theory (QLT, see Jokipii 1966),
\item Non-linear diffusion theory (see, e.g., Shalchi et al. 2004, Shalchi 2010);
\end{enumerate}
For both approaches we compute the perpendicular diffusion coefficient as a function of the gyroradius. Results
are compared with each other and previous findings are recovered by considering appropriate limits. Furthermore, we
estimate the value of $"a"$ in the context of finite gyroradius corrections. As examples we consider three different
turbulence models namely the slab model, a noisy slab model, and the two-dimensional model.

The remainder of this paper is organized as follows.  In Section 2 we briefly discuss the relation between particle and guiding
center coordinates and the corresponding diffusion coefficients. Quasi-linear theory is employed in Section 3 and in Section 4
we use a more advanced approach based on non-linear diffusion theory. We end with a short summary and some conclusions in Section 5.
\section{Relation between guiding center and particle coordinates}
Usually one is interested in the coordinates of the charged particle interacting with turbulence. In the following we refer
to these coordinates as {\it particle coordinates} and we will use the symbols $\vec{x}$ and $\vec{v}$ for particle position
and velocity, respectively. Alternatively, one can use the coordinates $\vec{R}$ defined via (see, e.g., Schlickeiser 2002)
\be
\vec{R} = \vec{x} + \frac{c}{q B_0} \left( \vec{p} \times \vec{e}_{z} \right)
= \vec{x} + \frac{1}{\Omega} \left( \vec{v} \times \vec{e}_{z} \right)
\label{eqBp1}
\ee
where $\Omega=(q B_0)/(m c \gamma)$ is the unperturbed gyrofrequency. Here we have used the electric charge of the particle
$q$, the rest mass $m$, the speed of light $c$, and the Lorentz factor $\gamma$. For the velocity $\vec{v}$ of the particle
we use spherical coordinates
\bdm
v_x & = & v \sqrt{1-\mu^2} \cos \Phi \nonumber\\
v_y & = & v \sqrt{1-\mu^2} \sin \Phi \nonumber\\
v_z & = & v \mu
\label{eq1p13}
\edm
with the particle speed $v$, the pitch-angle cosine $\mu$, and the gyrophase $\Phi$.

In the unperturbed case, where we have by definition $\delta \vec{B} = 0$, the vector $\vec{R}$ corresponds to the position
of the guiding center. Therefore, we call $\vec{R}$ the {\it guiding center coordinates}. To obtain the velocity of the
guiding center, we consider the time derivative of Eq. (\ref{eqBp1}). This gives us
\bdm
\vec{V} := \frac{d \vec{R}}{d t} & = & \vec{v} + \frac{c}{q B_0} \left( \frac{d \vec{p}}{d t} \times \vec{e}_{z} \right) \nonumber\\
& = & \vec{v} + \frac{1}{B_0} \left[ \left( \vec{v} \times \vec{B} \right) \times \vec{e}_z \right] \nonumber\\
& = & v_{z} \frac{\vec{B}}{B_0} - \vec{v} \frac{\delta B_{z}}{B_0}.
\label{eqBp2}
\edm
where we have employed the {\it Newton-Lorentz equation} $d \vec{p} / d t = q ( \vec{v} \times \vec{B} )/c$ and the {\it Gra\ss mann Identity}
\be
\vec{a} \times \left( \vec{b} \times \vec{c} \right) = \vec{b} \left( \vec{a} \cdot \vec{c} \right) - \vec{c} \left( \vec{a} \cdot \vec{b} \right).
\ee
Therefore, we can express the components $V_i$ of the guiding center velocity vector by the components of the particle
velocity vector $v_i$ and the magnetic field components. Very often in diffusion theory, turbulence models with $\delta B_z = 0$
are considered. Examples are the slab and the two-dimensional model (see below for a definition of these two models).
In this particular case Eq. (\ref{eqBp2}) simplifies to
\be
V_x = v_z \frac{\delta B_x}{B_0}, \quad V_y = v_z \frac{\delta B_y}{B_0}, \quad \textnormal{and} \quad V_z = v_z.
\label{GCvelo}
\ee
From a more practical point of view, these equations are valid as long as the condition $\delta B_z \ll B_0$ is satisfied.
In this case they can be used as starting point to compute the perpendicular diffusion coefficient.

We have to be very careful if {\it particle coordinates} or {\it guiding center coordinates} are used. The guiding center
coordinates ($X$, $Y$, $Z$) are related to particle coordinates ($x$, $y$, $z$) via
\be
X = x + \frac{v_y}{\Omega}, \quad Y = y - \frac{v_x}{\Omega}, \quad \textnormal{and} \quad Z = z
\label{relationxX}
\ee
as derived from Eq. (\ref{eqBp1}).

The perpendicular diffusion coefficients can be calculated as time derivatives of the corresponding mean square displacements.
From Eq. (\ref{relationxX}) we find the following relations
\bdm
\left< \left( \Delta X \right)^2 \right> & = & \left< \left( \Delta x \right)^2 \right> + \frac{2}{\Omega} \left< \Delta x \Delta v_y \right> \nonumber\\
& + & \frac{1}{\Omega^2} \left< \left( \Delta v_y \right)^2 \right>, \nonumber\\
\left< \left( \Delta Y \right)^2 \right> & = & \left< \left( \Delta y \right)^2 \right> - \frac{2}{\Omega} \left< \Delta y \Delta v_x \right> \nonumber\\
& + & \frac{1}{\Omega^2} \left< \left( \Delta v_x \right)^2 \right>;
\label{relmsd}
\edm
The mean square displacements $\langle \left( \Delta X \right)^2 \rangle$, $\langle \left( \Delta Y \right)^2 \rangle$,
$\langle \left( \Delta x \right)^2 \rangle$, and $\langle \left( \Delta y \right)^2 \rangle$ should increase linearly
with time if the transport is indeed diffusive. The terms $\left< \Delta x \Delta v_y \right>$ and $\left< \Delta y \Delta v_x \right>$
correspond basically to the drift coefficient (see, e.g., Shalchi 2011) and, thus, they should approach asymptotically a constant.
Since velocities are limited due to $|v_x| \leq v$ and $|v_y| \leq v$, the last terms in Eq. (\ref{relmsd}) can also not increase
with time. Therefore, we conclude that
\be
\langle \left( \Delta X \right)^2 \rangle = \langle \left( \Delta x \right)^2 \rangle
\quad \textnormal{and} \quad
\langle \left( \Delta Y \right)^2 \rangle = \langle \left( \Delta y \right)^2 \rangle
\ee
in the limit of late times. Thus the perpendicular diffusion coefficients of particles and guiding centers should be the same in that limit.
\section{Quasi-linear perpendicular diffusion}
As a first step we consider Quasi-Linear Theory (QLT) and take into account the gyrorotation of the particle. By doing this,
we basically follow Schlickeiser (2002). As explained in Shalchi (2015), QLT is correctly describing perpendicular diffusion
only for very long parallel mean free paths (corresponding to suppressed pitch-angle scattering) and small Kubo numbers.

For axis-symmetric turbulence, we can calculate the Fokker-Planck coefficient of perpendicular diffusion $D_{\perp}$ by using
the {\it Taylor-Green-Kubo formulation} (see Taylor 1922, Green 1951, Kubo 1957)
\be
D_{\perp} = \Re \int_{0}^{\infty} d t \; \left< V_x (t) V_x (0) \right>.
\label{usetgk}
\ee
By assuming $\delta B_z \ll B_0$, we can use the first formula of Eq. (\ref{GCvelo}) to derive
\be
D_{\perp} = \frac{1}{B_0^2} \Re \int_{0}^{\infty} d t \; \left< v_z (t) v_z (0) \delta B_x (t) \delta B_x (0) \right>
\label{eq4p1}
\ee
where $\delta B_x (t)$ stands for $\delta B_x [\vec{x}(t), t]$. Within the quasi-linear approximation we assume that pitch-angle
scattering is suppressed and, thus,
\be
v_z (t) v_z (0) = v^2 \mu^2 = constant.
\label{novelo}
\ee
Therewith, we derive
\be
D_{\perp} = \frac{v^2 \mu^2}{B_0^2} \Re \int_{0}^{\infty} d t \; \left< \delta B_x (t) \delta B_x (0) \right>.
\label{Dperpnovelo}
\ee
For the magnetic correlation function we can employ
\be
\left< \delta B_x (t) \delta B_x (0) \right> = \int d^3k \; P_{xx} (\vec{k})
e^{i \vec{k} \cdot \left[ \vec{x}_u (t) - \vec{x}_u (0) \right]}.
\label{introPxx}
\ee
Here we have used the magnetic correlation tensor in the Fourier space $P_{lm} (\vec{k}) = \langle \delta B_l (\vec{k}) \delta B_m^* (\vec{k}) \rangle$
and $\vec{x}_u (t)$ denotes the unperturbed orbit. Furthermore, we have assumed that the turbulence is static. In the following, we use cylindrical
coordinates for the wavevector $\vec{k}$ which are related to Cartesian coordinates via
\bdm
k_{\parallel} & = & k_{z}, \nonumber\\
k_{\perp} & = & \sqrt{k_{x}^2+k_{y}^2}, \nonumber\\
\Psi & = & \textnormal{arccot}(k_{x} / k_{y});
\label{eq2p6}
\edm
According to Eq. (12.2.2a) of Schlickeiser (2002), we have within QLT
\be
e^{i \vec{k} \cdot \vec{x}_u (t)} = \sum_{n=-\infty}^{+\infty} J_n (W)
e^{i v \mu k_{\parallel} t + i n \left( \Psi - \Phi_0 + \Omega t \right)}
\label{expoqlt}
\ee
with
\be
W = R_L k_{\perp} \sqrt{1 - \mu^2}
\ee
and the initial gyrophase $\Phi_0$. The parameter $R_L$ denotes the unperturbed Larmor radius / gyroradius at $\mu = 0$.
In Eq. (\ref{expoqlt}) we have also used the Bessel function $J_n (W)$. At the initial time $t=0$, Eq. (\ref{expoqlt}) becomes
\be
e^{i \vec{k} \cdot \vec{x}_u (0)} = \sum_{m=-\infty}^{+\infty} J_m (W) e^{i m \left( \Psi - \Phi_0 \right)}.
\label{qltexp}
\ee
By combining Eqs. (\ref{expoqlt}) and (\ref{qltexp}), by averaging over the initial gyrophase, and by employing
\be
\int_{0}^{2 \pi} d \Phi_0 \; e^{i \Phi_0 \left( m - n \right)} = 2 \pi \delta_{nm},
\label{gyrodelta}
\ee
we derive
\be
e^{i \vec{k} \cdot \left[ \vec{x}_u (t) - \vec{x}_u (0) \right]} = \sum_{n=-\infty}^{+\infty} J_n^2 (W) e^{i \left( v \mu k_{\parallel} + n \Omega \right) t}.
\label{qltexpodiff}
\ee
With Eq. (\ref{qltexpodiff}) we can write Eq. (\ref{introPxx}) as
\bdm
& & \left< \delta B_x (t) \delta B_x (0) \right> \nonumber\\
& = & \int d^3k \; P_{xx} (\vec{k}) \nonumber\\
& \times & \sum_{n=-\infty}^{+\infty} J_n^2 (W) e^{i \left( v \mu k_{\parallel} + n \Omega \right) t}
\edm
and the Fokker-Planck coefficient of perpendicular diffusion (\ref{Dperpnovelo}) can be written as
\bdm
D_{\perp} & = & \pi \left( \frac{v \mu}{B_0} \right)^2 \int d^3k \; P_{xx} (\vec{k}) \nonumber\\
& \times & \sum_{n=-\infty}^{+\infty} J_n^2 (W) \delta \left( v \mu k_{\parallel} + n \Omega \right)
\label{QLT}
\edm
where we have used the relation (see, e.g., Hoskins 2009)
\be
\int_{0}^{\infty} d t \; e^{i x t} = \pi \delta \left( x \right)
\ee
with Dirac's delta distribution $\delta (x)$. In order to evaluate Eq. (\ref{QLT}), one has to specify the tensor
component $P_{xx}$ or consider special limits. This is done in the following paragraphs.
\subsection{Zero gyroradius limit}
Eq. (\ref{QLT}) depends on the unperturbed gyroradius $R_L$ due to $W=R_L k_{\perp} \sqrt{1 - \mu^2}$ in the Bessel functions.
If we consider the limit $R_L \rightarrow 0$, and therewith $W \rightarrow 0$, we can use (see, e.g., Abramovitz \& Stegun 1974)
\be
J_n^2 (0) = \delta_{n0},
\label{besselid}
\ee
and Eq. (\ref{QLT}) becomes
\be
D_{\perp} = \pi \left( \frac{v \mu}{B_0} \right)^2 \int d^3k \;
P_{xx} (\vec{k}) \delta \left( v \mu k_{\parallel} \right).
\ee
Now we employ the relation (see, e.g., Hoskins 2009)
\be
\delta \left( a x \right) = \frac{1}{\left| a \right|} \delta \left( x \right)
\label{diracid}
\ee
to obtain
\be
D_{\perp} = \pi \frac{v \left| \mu \right|}{B_0^2} \int d^3k \;
P_{xx} (\vec{k}) \delta \left( k_{\parallel} \right).
\ee
In the latter formula we find the parallel integral scale $L_{\parallel}$ which is defined via (see, e.g., Shalchi 2014)
\be
\delta B_x^2 L_{\parallel} = \pi \int d^3 k \; P_{xx} \left( \vec{k} \right) \delta \left( k_{\parallel} \right).
\label{Lparallel}
\ee
By using $L_{\parallel}$, we can write the Fokker-Planck coefficient of perpendicular diffusion as
\be
D_{\perp} = \frac{1}{2} v \left| \mu \right| L_{\parallel} \frac{\delta B^2}{B_0^2}.
\label{Dperpzero}
\ee
By employing (see, e.g., Schlickeiser 2002)
\be
\kappa_{\perp} = \frac{1}{2} \int_{-1}^{+1} d \mu \; D_{\perp} \left( \mu \right),
\label{kappaandDperp}
\ee
we can easily calculate the perpendicular diffusion coefficient $\kappa_{\perp}$ and we find the well-known
quasi-linear result
\be
\kappa_{\perp} = \frac{1}{4} v L_{\parallel} \frac{\delta B^2}{B_0^2}.
\label{kappaperpzero}
\ee
We would like to emphasize that the latter formula is correct within the quasi-linear approximation and if we consider
the limit $R_L \rightarrow 0$. Eq. (\ref{kappaperpzero}) is usually called the (quasi-linear) field line random walk
limit and was originally obtained in Jokipii (1966).
\subsection{Slab turbulence}
In the following we consider the so-called slab model for which the turbulent magnetic field depends only on the coordinate along
the guide field, i.e., $\delta \vec{B} (\vec{x}) = \delta \vec{B} (z)$. In this particular case the components of the magnetic
correlation tensor are given by
\be
P_{lm} (\vec{k}) = g^{slab} (k_{\parallel}) \frac{\delta (k_{\perp})}{k_{\perp}} \delta_{lm}.
\label{slabmode}
\ee
Here we have used {\it Kronecker's delta} $\delta_{lm}$, {\it Dirac's delta distribution} $\delta (z)$, and $l,m=x,y$.
To satisfy the solenoidal constraint, we need to have $P_{lz}=P_{zm}=P_{zz}=0$. Here $g^{slab} (k_{\parallel})$
is the turbulence spectrum of the slab modes.

If we combine Eqs. (\ref{QLT}) and (\ref{slabmode}) we find
\bdm
D_{\perp} & = & 2 \pi^2 \left( \frac{v \mu}{B_0} \right)^2 \int_{-\infty}^{+\infty} d k_{\parallel} \; g^{slab} (k_{\parallel}) \nonumber\\
& \times & \sum_{n=-\infty}^{+\infty} J_n^2 (0) \delta \left( v \mu k_{\parallel} + n \Omega \right).
\edm
Employing again Eq. (\ref{besselid}) is leading to
\be
D_{\perp} = 2 \pi^2 \left( \frac{v \mu}{B_0} \right)^2
\int_{-\infty}^{+\infty} d k_{\parallel} \; g^{slab} (k_{\parallel}) \delta \left( v \mu k_{\parallel} \right).
\ee
Now we use Eq. (\ref{diracid}) to obtain
\bdm
D_{\perp} & = & 2 \pi^2 \frac{v \left| \mu \right|}{B_0^2} \int_{-\infty}^{+\infty} d k_{\parallel} \; g^{slab} (k_{\parallel}) \delta \left( k_{\parallel} \right) \nonumber\\
& = & 2 \pi^2 \frac{v \left| \mu \right|}{B_0^2} g^{slab} (k_{\parallel} = 0).
\label{Dperpslabspec1}
\edm
To continue, we need to specify the function $g^{slab} (k_{\parallel})$ in Eq. (\ref{Dperpslabspec1}). In the following we employ 
\be
g^{slab} (k_{\parallel}) = \frac{C(s)}{2 \pi} \delta B^2 l_{\parallel} \left[ 1 + (k_{\parallel} l_{\parallel})^2 \right]^{-s/2}
\label{slabspec}
\ee
as suggested by Bieber et al. (1994). Here we have used the strength of the turbulent field $\delta B$ and the bendover scale $l_{\parallel}$
indicating the turnover from the energy range of the spectrum to the inertial range. The normalization function $C(s) = \Gamma (s/2) / [2 \sqrt{\pi} \Gamma ((s-1)/2)]$
depends on the inertial range spectral index $s$ and contains Gamma functions. Therewith, Eq. (\ref{Dperpslabspec1}) becomes
\be
D_{\perp} = \pi C (s) v \left| \mu \right| l_{\parallel} \frac{\delta B^2}{B_0^2}.
\label{Dperpslabspec2}
\ee
One can easily show that for the spectrum used here, the two parallel length scales are related to each other via
$L_{\parallel} = 2 \pi C (s) l_{\parallel}$. Therefore, Eq. (\ref{Dperpslabspec2}) agrees with Eq. (\ref{Dperpzero})
and, thus, Eq. (\ref{kappaperpzero}) is also valid in the case considered here.

We have shown that for slab turbulence, the zero gyroradius limit is exact if quasi-linear theory is employed. This result
was predictable because for slab turbulence we have by definition $\delta B_x = \delta B_x (z)$ and, thus, the perpendicular
motion of the particle is not relevant.
\subsection{A noisy slab model}
The slab model used in the previous paragraph is an extreme model due to the fact that there is absolutely no transverse structure
in the turbulence. Therefore, we extend the slab model by employing
\bdm
P_{lm} \left( \vec{k} \right) & = & \frac{2 l_{\perp}}{k_{\perp}} g^{slab} (k_{\parallel}) \Theta \left( 1 - k_{\perp} l_{\perp} \right) \nonumber\\
& \times & \left( \delta_{lm} - \frac{k_{l} k_{m}}{k_{\perp}^2} \right)
\label{noisyslab}
\edm
where we have used the {\it Heaviside step function} $\Theta (x)$. We refer to this model as the {\it noisy slab model}
which was originally described in Shalchi (2015). Due to the step function, there is only turbulence if
$k_{\perp} l_{\perp} \leq 1$. For large enough wavenumbers, on the other hand, there is no turbulence. Physically this
corresponds to the case where wave vectors are mainly oriented parallel with respect to the mean field but weak fluctuations
are taken into account. The idea of incorporating some kind of noisiness via a step function goes back to Ruffolo \& Matthaeus (2013)
where this idea was combined with the two-dimensional turbulence model. A more detailed explanation of this type of turbulence
model can be found there.

Combining the noisy slab model represented by Eq. (\ref{noisyslab}) with Eq. (\ref{QLT}) yields
\bdm
D_{\perp}
& = & 2 \pi^2 \left( \frac{v \mu}{B_0} \right)^2 l_{\perp} \int_{0}^{l_{\perp}^{-1}} d k_{\perp} \; \sum_{n=-\infty}^{+\infty} J_n^2 (W) \nonumber\\
& \times & \int_{-\infty}^{+\infty} d k_{\parallel} \; g^{slab} (k_{\parallel}) \delta \left( v \mu k_{\parallel} + n \Omega \right).
\edm
Now we employ Eq. (\ref{diracid}) and assume that the spectrum is symmetric, i.e., $g^{slab} (k_{\parallel}) = g^{slab} (-k_{\parallel})$.
After evaluating the $k_{\parallel}$-integral we obtain
\bdm
D_{\perp} & = & 2 \pi^2 l_{\perp} \frac{v \left| \mu \right|}{B_0^2} \int_{0}^{l_{\perp}^{-1}} d k_{\perp} \; \nonumber\\
& \times & \sum_{n=-\infty}^{+\infty} J_n^2 (W) g^{slab} \left( k_{\parallel} = \frac{n \Omega}{v \mu} \right).
\edm
To proceed we employ spectrum (\ref{slabspec}) to deduce
\bdm
D_{\perp} & = & \pi C(s) l_{\parallel} l_{\perp} \frac{\delta B^2}{B_0^2} v \left| \mu \right| \left| \mu R \right|^{s} \nonumber\\
& \times & \int_{0}^{l_{\perp}^{-1}} d k_{\perp} \sum_{n=-\infty}^{+\infty} \frac{J_n^2 (W)}{\left[ n^2 + \left( \mu R \right)^2 \right]^{s/2}}
\label{noisywithKepteyn}
\edm
where we have used $R = R_L / l_{\parallel} = v /(\Omega l_{\parallel})$. We can easily recover the pure slab result (\ref{Dperpslabspec2})
by considering the limit $l_{\perp} \rightarrow \infty$.

To achieve a further simplification of Eq. (\ref{noisywithKepteyn}), we consider the special case $s=2$. In this case we can
use $C (s=2) = 1/(2 \pi)$ to write
\bdm
D_{\perp} & = & \frac{1}{2} l_{\parallel} l_{\perp} \frac{\delta B^2}{B_0^2} v \left| \mu \right| \left( \mu R \right)^{2} \nonumber\\
& \times & \int_{0}^{l_{\perp}^{-1}} d k_{\perp} \; \sum_{n=-\infty}^{+\infty} \frac{J_n^2 (W)}{n^2 + \left( \mu R \right)^2}.
\label{noisyperp1}
\edm
We would like to add, that in this case the parallel integral scale $L_{\parallel}$ and the bendover scale $l_{\parallel}$ are equal.
In order to evaluate the form (\ref{noisyperp1}) analytically, one has to approximate the series involving Bessel functions. The latter series
is discussed in detail in the Appendix of the current paper. These discussions are based on the work of Shalchi \& Schlickeiser (2004)
and Tautz \& Lerche (2010), respectively. 

\subsection{Noisy slab turbulence and small gyroradii}
In the following we focus on the case of small gyroradius corrections. Therefore, we assume that the parameters $W=R_L k_{\perp} \sqrt{1-\mu^2}$
and $\mu R$ are small as well. In this particular case we can employ Eq. (\ref{Tautzlimit2}). Therewith, Eq. (\ref{noisyperp1}) can be
approximated by
\bdm
D_{\perp} & \approx & \frac{1}{2} l_{\parallel} l_{\perp} \frac{\delta B^2}{B_0^2} v \left| \mu \right| \nonumber\\
& \times & \int_{0}^{l_{\perp}^{-1}} d k_{\perp} \left[ 1 - \frac{1}{2} \left( 1 - \mu^2 \right) R_L^2 k_{\perp}^2 \right].
\edm
The remaining wavenumber integrals can easily be solved and we deduce
\be
D_{\perp} \approx \frac{1}{2} l_{\parallel} \frac{\delta B^2}{B_0^2} v \left| \mu \right|
\left[ 1 - \frac{1}{6} \left( 1 - \mu^2 \right) \frac{R_L^2}{l_{\perp}^2} \right].
\ee
Employing Eq. (\ref{kappaandDperp}) allows as to compute the perpendicular diffusion coefficient. After straightforward algebra we find
\be
\kappa_{\perp} \approx \frac{1}{4} v l_{\parallel} \frac{\delta B^2}{B_0^2} \left( 1 - \frac{1}{12} \frac{l_{\parallel}^2}{l_{\perp}^2} R^2 \right)
\label{redkappasmall}
\ee
where we have used again $R = R_L / l_{\parallel}$. Or, if one is interested in the perpendicular mean free path
\be
\lambda_{\perp} \approx \frac{3}{4} l_{\parallel} \frac{\delta B^2}{B_0^2} \left( 1 - \frac{1}{12} \frac{l_{\parallel}^2}{l_{\perp}^2} R^2 \right).
\label{redlambdasmall}
\ee
If the parameter $"a^2"$ used in Eq. (\ref{introa}) is understood as finite Larmor radius corrections, we find for the case considered here
\be
a^2 \approx 1 - \frac{1}{12} \frac{l_{\parallel}^2}{l_{\perp}^2} R^2.
\ee
Here we can directly see, that finite gyroradius corrections depend on the normalized gyroradius $R$ and the scale ratio $l_{\parallel}/l_{\perp}$.
This conclusion is in agreement with Fig. \ref{NoisyQLT}. Furthermore, we conclude that the parameter $"a^2"$ is smaller than one.
\subsection{Numerical results for noisy slab turbulence}
\begin{figure}
\centering 
\includegraphics[width=0.50\textwidth]{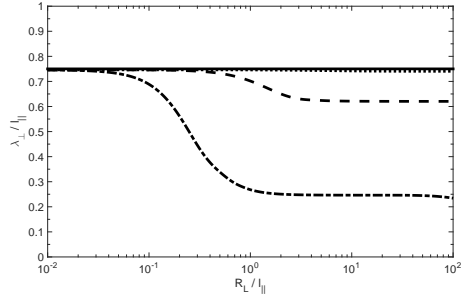}
\caption{The perpendicular mean free path $\lambda_{\perp}$ versus the unperturbed gyroradius $R_L$. Both parameters are normalized
with respect to the parallel scale $l_{\parallel}$. All curves shown here are obtained for quasi-linear transport in noisy
slab turbulence. The solid line corresponds to the zero Larmor radius limit and the other curves are obtained by taking into account
the non-vanishing gyroradius. Shown are the results for $l_{\parallel} / l_{\perp} = 0.1$ (dotted line), $l_{\parallel} / l_{\perp} = 1$
(dashed line), and $l_{\parallel} / l_{\perp} = 10$ (dash-dotted line).}
\label{NoisyQLT}
\end{figure}

For arbitrary Larmor radii $R_L$, we can compute the perpendicular mean free path only by solving Eq. (\ref{noisyperp1}) numerically.
The results are visualized in Fig. \ref{NoisyQLT}. There, we have also shown the zero Larmor-radius limit. The latter limit can be
obtained from Eq. (\ref{redlambdasmall}) by setting $R=0$ therein. In this particular case the perpendicular mean free path becomes
\be
\frac{\lambda_{\perp}}{l_{\parallel}} = \frac{3}{4} \frac{\delta B^2}{B_0^2}.
\ee
Furthermore, we have computed the ratio $\lambda_{\perp}/l_{\parallel}$ versus $R_L/l_{\parallel}$ for different values of the
ratio $l_{\parallel}/l_{\perp}$. In all considered cases we find that finite Larmor-radius corrections reduce the perpendicular
diffusion coefficient. For $l_{\perp} = 0.1 l_{\parallel}$, for instance, the perpendicular mean free path is reduced by a factor $3$.
\subsection{Two-dimensional turbulence}
Another popular model for turbulence is the two-dimensional (2D) model (see, e.g., Shalchi 2009 for details). In this case the
magnetic correlation tensor is given by
\be
P_{lm} (\vec{k}) = g^{2D}(k_{\perp}) \frac{\delta (k_{\parallel})}{k_{\perp}} \left( \delta_{lm} - \frac{k_{l} k_{m}}{k_{\perp}^2} \right)
\label{eq5p1c}
\ee
if $l,m=x,y$. The tensor used here corresponds to the one defined in Bieber et al. (1994). In the latter paper it was also assumed
that $\delta B_z = 0$. Therefore, tensor components with $l=z$ and/or $m=z$ are zero. The function $g^{2D}(k_{\perp})$ used above
is the spectrum of the two-dimensional modes.

By combining Eq. (\ref{eq5p1c}) with Eq. (\ref{QLT}) we find after straightforward algebra
\bdm
D_{\perp} & = & \pi^2 \left( \frac{v \mu}{B_0} \right)^2 \int_{0}^{\infty} d k_{\perp} \; g^{2D}(k_{\perp}) \nonumber\\
& \times & \sum_{n=-\infty}^{+\infty} J_n^2 (W) \delta \left( n \Omega \right).
\label{QLT2D}
\edm
Only the contribution with $n=0$ is not zero due to the Dirac delta $\delta (n \Omega)$. For $n=0$, however, we have $\delta (0) = \infty$
and, thus, $D_{\perp} = \infty$. Therefore, QLT provides for the two-dimensional turbulence model an infinite perpendicular diffusion coefficient.
This result is unphysical and was found before in diffusion theory (see, e.g., Shalchi et al. 2004). Therefore, QLT cannot be used in order to
describe perpendicular transport in the general case. In the next section we consider transport in two-dimensional turbulence again but in the
context of non-linear diffusion theory.
\section{Non-linear diffusion theory}
It is well-known and accepted that quasi-linear theory cannot describe perpendicular diffusion in the general case (see, e.g.,
Spatschek 2008 and Shalchi 2009). Even if one would assume that pitch-angle scattering is suppressed and perpendicular diffusion
is only caused due to the stochasticity of magnetic field lines, quasi-linear theory would not work due to the fact that
the field line random walk itself is non-linear for large Kubo numbers (see, e.g., Shalchi 2015). Therefore, a non-linear treatment
is required in order to describe perpendicular transport. A very powerful tool is the Unified-Non-Linear Transport (UNLT)
theory of Shalchi (2010). The latter theory can describe transport of magnetic field lines and particles across the mean field.
The theories of Matthaeus et al. (1995) and Matthaeus et al. (2003) as well as quasi-linear theory are contained in this theory
as special limits.

In the following we use ideas of non-linear diffusion theory for the special case of suppressed pitch-angle scattering. The case
of strong pitch-angle scattering will be investigated in the sequel of the current paper. According to Eq. (\ref{eq4p1}) the critical
quantity in the theory of perpendicular diffusion is the 4th order correlation function
\be
C (t_1, t_2) := \left< v_z (t_1) v_z (t_2) \delta B_x (t_1) \delta B_x (t_2) \right>
\label{defineC}
\ee
where we have used again the notation $\delta B_x (t) = \delta B_x [\vec{x} (t),t]$. Such 4th order correlations are difficult
to evaluate. Matthaeus et al. (2003), for instance, suggested to approximate 4th order correlations by a product of two 2nd order
correlations. In the general case, however, this approximation is not valid (see Shalchi 2010). In the current paper we consider
only the case that velocity diffusion is suppressed and, thus, we can approximate
\bdm
& & \left< v_z (t) v_z (0) \delta B_x (t) \delta B_x (0) \right> \nonumber\\
& \approx & v^2 \mu^2 \int d^3 k \; P_{xx} (\vec{k}) \left< e^{i \vec{k} \cdot \left[ \vec{x} (t) - \vec{x} (0) \right]} \right>.
\label{usedcorr}
\edm
Here we have also employed the so-called {\it random-phase approximation} which is a well-known tool in diffusion theory
(see, e.g., Lerche 1973, Matthaeus et al. 1995, Matthaeus et al. 2003, Tautz \& Shalchi 2010). The latter approximation
is sometimes called {\it Corrsin's independence hypothesis} (see Corrsin 1959).

Here one has to be careful since the magnetic field has to be computed at the particle position. Therefore, $\vec{x} (t)$
and $\vec{x} (0)$ are particle positions and not guiding center positions. To evaluate the characteristic function
$\langle \exp{(i \vec{k} \cdot \vec{x})} \rangle$, previous authors used the propagator of the diffusion equation (see Matthaeus et al. 2003)
or an approach based on the Fokker-Planck equation (see Shalchi 2010). However, the latter two equations are derived
by using guiding center coordinates (see, e.g., Schlickeiser 2002).

With Eq. (\ref{relationxX}), the characteristic function in Eq. (\ref{usedcorr}) becomes
\bdm
\left< e^{i \vec{k} \cdot \left[ \vec{x} (t) - \vec{x} (0) \right]} \right> & = & \left< e^{i \vec{k} \cdot \left[ \vec{X} (t) - \vec{X} (0) \right]} \right> \nonumber\\
& \times & e^{i k_x \Delta v_x / \Omega - i k_x \Delta v_y / \Omega}.
\edm
Alternatively, one can write
\be
\left< e^{i \vec{k} \cdot \left[ \vec{x} (t) - \vec{x} (0) \right]} \right>
= e^{i \vec{k} \cdot \left[ \vec{x}_{u} (t) - \vec{x}_u (0) \right]} \left< e^{i \Delta X k_x + i \Delta Y k_y} \right>
\label{exprelations}
\ee
where we have used the unperturbed orbit $x_u (t)$ as in Section 3.

The UNLT theory allows in principle to calculate $D_{\perp}$ non-linearly. By using the latter approach, 4th order correlations
of the form $\langle v_x (t) v_x (0) \delta B_x (t) \delta B_x (0) \rangle$ can be computed by using the Fokker-Planck equation
which has the form (see, e.g., Schlickeiser 2002)
\bdm
& & \frac{\partial f}{\partial t} + v \mu \frac{\partial f}{\partial Z} - \Omega \frac{\partial f}{\partial \Phi} \nonumber\\
& = & \frac{\partial}{\partial \mu} \left( D_{\mu\mu} \frac{\partial f}{\partial \mu} \right)
+ \frac{\partial}{\partial \Phi} \left( D_{\Phi\Phi} \frac{\partial f}{\partial \Phi} \right) \nonumber\\
& + & D_{\perp} \left( \frac{\partial^2 f}{\partial X^2} + \frac{\partial^2 f}{\partial Y^2} \right).
\label{thefpeq}
\edm
Here we have assumed axis-symmetry with respect to the $z$-axis and we have used the Fokker-Planck coefficients
of pitch-angle diffusion $D_{\mu\mu}$ and gyrophase diffusion $D_{\Phi\Phi}$, respectively. In non-linear diffusion theories
a fundamental quantity is the so-called {\it characteristic function} which is defined as
\be
\Gamma (\vec{k},\mu,t) := \int d^3 X \; e^{i \vec{k} \cdot \vec{X}} f (\vec{X}, \mu, \mu_0, t) \equiv \left< e^{i \vec{k} \cdot \vec{X}} \right>.
\label{defgamma}
\ee
To proceed we multiply the Fokker-Planck equation (\ref{thefpeq}) by $\exp{(i \vec{k} \cdot \vec{X})}$ and thereafter we integrate over space to obtain
\bdm
& & \frac{\partial \Gamma}{\partial t} - i k_{z} v \mu \Gamma - \Omega \frac{\partial f}{\partial \Phi} \nonumber\\
& = & \frac{\partial}{\partial \mu} \left[ D_{\mu\mu} \frac{\partial \Gamma}{\partial \mu} \right]
+ \frac{\partial}{\partial \Phi} \left[ D_{\Phi\Phi} \frac{\partial \Gamma}{\partial \Phi} \right]
- D_{\perp} k_{\perp}^2 \Gamma.
\label{ODEforGamma}
\edm
If we suppress pitch-angle scattering and gyrophase diffusion by setting $D_{\mu\mu} = 0$ and $D_{\Phi\Phi} = 0$, the latter
equation is solved by
\be
\Gamma (\vec{k},\mu,t) = e^{i k_{z} v \mu t - D_{\perp} k_{\perp}^2 t}
\label{wnltgamma}
\ee
where we have also average over the gyrophase $\Phi$. This formula provides the characteristic function for suppressed velocity diffusion.
Therefore, Eq. (\ref{exprelations}) becomes in the case considered here
\be
\left< e^{i \vec{k} \cdot \left[ \vec{x} (t) - \vec{x} (0) \right]} \right>
= e^{i \vec{k} \cdot \left[ \vec{x}_{u} (t) - \vec{x}_u (0) \right]} e^{- D_{\perp} k_{\perp}^2 t}
\label{exprelations2}
\ee
where $\vec{x}_{u} (t)$ denotes the unperturbed orbit as before.

To proceed we combine Eqs. (\ref{eq4p1}), (\ref{qltexpodiff}), (\ref{usedcorr}), and (\ref{exprelations2}) to derive for the
Fokker-Planck coefficient of perpendicular diffusion
\be
D_{\perp} = \frac{v^2 \mu^2}{B_0^2} \sum_{n=-\infty}^{+\infty} \int d^3 k \; R_{n} (\vec{k}) P_{xx} (\vec{k}) J_n^2(W)
\label{eq4p3}
\ee
with the resonance function
\bdm
R_{n} (\vec{k}) & = & Re \int_{0}^{\infty} dt \; e^{i (v \mu k_{\parallel} + n \Omega) t} e^{-D_{\perp} k_{\perp}^{2} t} \nonumber\\
& = & \frac{D_{\perp} k_{\perp}^2}{\left( D_{\perp} k_{\perp}^2 \right)^2 + \left( v \mu k_{\parallel} + n \Omega \right)^2}.
\label{eq3p14}
\edm
We can easily see that perpendicular diffusion described by the parameter $D_{\perp}$ broadens the resonance. The effect of resonance
broadening due to perpendicular diffusion was already described in Shalchi et al. (2004).

The quasi-linear limit (\ref{QLT}) can be recovered by considering $D_{\perp} \rightarrow 0$ and by using (see, e.g., Hoskins 2009)
\be
\lim_{a \rightarrow 0} \frac{a}{a^2 + x^2} \rightarrow \pi \delta \left( x \right).
\label{breitwigner}
\ee
This is leading to the results discussed in Section 3 of the present paper.

In the current section, Eq. (\ref{eq4p3}) with (\ref{eq3p14}) was derived by using the approach proposed in Shalchi (2010) for a finite
gyroradius but suppressed velocity diffusion. We would like to emphasize that the same result can be obtained from the so-called
{\it Weakly Non-Linear Theory (WNLT)} of Shalchi et al. (2004) if velocity diffusion is suppressed therein.
\subsection{Zero gyroradius limit}
Eq. (\ref{eq4p3}) with (\ref{eq3p14}) describes perpendicular transport for suppressed velocity diffusion but with finite gyorradius
in the non-linear case. We can recover the zero gyroradius limit by considering $W \rightarrow 0$. In this case we can use again
Eq. (\ref{besselid}) and Eq. (\ref{eq4p3}) becomes
\be
D_{\perp} = \frac{v^2 \mu^2}{B_0^2} \int d^3 k \; R_{0} (\vec{k}) P_{xx} (\vec{k})
\label{unlt0}
\ee
with
\be
R_{0} (\vec{k}) = \frac{D_{\perp} k_{\perp}^2}{\left( D_{\perp} k_{\perp}^2 \right)^2 + \left( v \mu k_{\parallel} \right)^2}.
\label{R0}
\ee
This result agrees with the UNLT theory for the case of suppressed pitch-angle scattering. As already discussed in Shalchi (2010),
Eq. (\ref{unlt0}) with (\ref{R0}) can be solved by
\be
D_{\perp} (\mu) = v \left| \mu \right| \kappa_{FL}
\label{FLRWlimit}
\ee
where $\kappa_{FL}$ is the solution of the following non-linear integral equation
\be
\kappa_{FL} = \frac{1}{B_0^2} \int d^3 k \; P_{xx} \left( \vec{k} \right)
\frac{\kappa_{FL} k_{\perp}^2}{\left( \kappa_{FL} k_{\perp}^2 \right)^2 + k_{\parallel}^2}.
\label{Bill95}
\ee
Eq. (\ref{FLRWlimit}) corresponds to the {\it Field Line Random Walk (FLRW) limit} and the parameter $\kappa_{FL}$ is the diffusion
coefficient of field line wandering. Eq. (\ref{Bill95}) is a non-linear integral equation for $\kappa_{FL}$ which was originally
derived by Matthaeus et al. (1995).
\subsection{Slab turbulence}
In the following we employ again the slab model defined via Eq. (\ref{slabmode}). If the latter model is combined with
Eq. (\ref{eq4p3}), we deduce
\bdm
D_{\perp} & = & 2 \pi \frac{v^2 \mu^2}{B_0^2} \sum_{n=-\infty}^{+\infty} J_n^2(0) \nonumber\\
& \times & \int_{-\infty}^{+\infty} d k_{\parallel} \; g^{slab} (k_{\parallel}) R_{n} (k_{\parallel},k_{\perp}=0).\nonumber\\
\edm
To proceed we use again Eq. (\ref{besselid}) to obtain
\be
D_{\perp} = 2 \pi \frac{v^2 \mu^2}{B_0^2} \int_{-\infty}^{+\infty} d k_{\parallel} \; g^{slab} (k_{\parallel}) R_{0} (k_{\parallel},k_{\perp}=0).
\ee
Now we combine Eq. (\ref{R0}) with (\ref{breitwigner}), and we employ again Eq. (\ref{diracid}) to find
\be
D_{\perp} = 2 \pi^2 \frac{v \left| \mu \right|}{B_0^2} \int_{-\infty}^{+\infty} d k_{\parallel} \; g^{slab} (k_{\parallel}) \delta (k_{\parallel})
\ee
which agrees with the slab result derived above. Obviously, finite Larmor radius corrections are not important in this case.
This result was predictable, because in the current section we only included the factor $\Gamma (k_{\parallel} = 0) = \exp (-D_{\perp} k_{\perp}^2 t)$
to the quasi-linear formula. For slab turbulence this factor is one.
\subsection{Two-dimensional turbulence}
As shown above, the non-linear effect due to perpendicular diffusion is not relevant for slab turbulence. Therefore, we have to consider
a model with transverse structure. A simple model, for which non-linear effects are important, is the two-dimensional model used already
above in the context of QLT. For this turbulence model we can employ Eq. (\ref{eq5p1c}) and therewith Eq. (\ref{eq4p3}) becomes
\bdm
D_{\perp} & = & \pi \frac{v^2 \mu^2}{B_0^2} \sum_{n=-\infty}^{+\infty} \int_{0}^{\infty} d k_{\perp} \; g^{2D}(k_{\perp}) \nonumber\\
& \times & R_{n} \left( k_{\parallel}=0, k_{\perp} \right) J_n^2(W)
\edm
with 
\be
R_{n} \left( k_{\parallel}=0, k_{\perp} \right)
= \frac{D_{\perp} k_{\perp}^2}{\left( D_{\perp} k_{\perp}^2 \right)^2 + \left( n \Omega \right)^2}.
\ee
As shown in Shalchi et al. (2004), the latter formulas  for suppressed velocity diffusion can be written as
\be
D_{\perp} = \frac{\pi v^2 \mu^2}{\Omega B_0^2} \int_{0}^{\infty} d k_{\perp} \; g^{2D}(k_{\perp}) V H(V,W)
\label{eq5p20}
\ee
where we have used the series
\be
H(V,W) = \sum_{n=-\infty}^{+\infty} \frac{J_{n}^2 (W)}{V^2 + n^2}
\label{eq5p21}
\ee
and the parameter $V = D_{\perp} k_{\perp}^2 / \Omega$. The series (\ref{eq5p21}) occurred already above in the context
of QLT (see Eq. (\ref{noisyperp1})). A more detailed discussion of this series can be found in the Appendix of the current article.

For the spectrum of the two-dimensional modes, we employ the model proposed by Shalchi \& Weinhorst (2009)
\bdm
g^{2D}(k_{\perp}) & = & \frac{2 D(s,q)}{\pi} \delta B^2 l_{\perp} \nonumber\\
& \times & \frac{(k_{\perp} l_{\perp})^{q}}{\left[ 1 + (k_{\perp} l_{\perp})^2 \right]^{(s+q)/2}}.
\label{2dspec}
\edm
Here we have used the perpendicular bendover scale $l_{\perp}$, the inertial range spectral index $s$, and the energy range spectral index $q$.
For the inertial range spectral index we employ $s=5/3$ as suggested by Kolmogorov (1941). The physical consequences of the different values
of $q$ are discussed in Matthaeus et al. (2007) and Shalchi \& Weinhorst (2009). It was shown there that in order to obtain a finite ultra-scale,
the condition $q>1$ must be satisfied. In the current paper we set $q=3$ as an example. In the spectrum (\ref{2dspec}) we have used the
normalization function
\be
D(s, q) = \frac{\Gamma \left( \frac{s+q}{2} \right)}{2 \Gamma \left( \frac{s-1}{2} \right) \Gamma \left( \frac{q+1}{2} \right)}
\label{normalD}
\ee
with the Gamma function $\Gamma (z)$.
\subsection{Two-dimensional turbulence and small gyroradii}
In the following we evaluate Eq. (\ref{eq5p20}) in the limit of small gyroradii. We do this by considering the asymptotic limit
$R_L \rightarrow 0$. Therefore, the parameter $W$, which is directly proportional to $R_L$, can be assumed to be small. More complicated
to estimate is the parameter $V$ which can be written as
\be
V = \frac{D_{\perp} k_{\perp}^2}{\Omega} \approx \frac{\kappa_{\perp} k_{\perp}^2}{\Omega}
= \frac{1}{3} \frac{R_L}{l_{\perp}} \frac{\lambda_{\perp}}{l_{\perp}} \left( l_{\perp} k_{\perp} \right)^2.
\ee
Since we consider the limit $R_L \rightarrow 0$, we can also assume that $V$ is small, at least as long as the perpendicular mean
free path is smaller than the perpendicular scale $l_{\perp}$. The latter condition is usually satisfied (see, e.g., Fig. \ref{2DNLT}
of the current paper). Thus, we have to approximate the series (\ref{eq5p21}) for the case of small values of $V$ and $W$. In this
special case we can employ approximation (\ref{Tautzlimit2}) and Eq. (\ref{eq5p20}) becomes
\bdm
D_{\perp}^2 & \approx & \pi \left( \frac{v \mu}{B_0} \right)^2 \int_{0}^{\infty} d k_{\perp} \; g^{2D}(k_{\perp}) k_{\perp}^{-2} \nonumber\\
& \times & \left[ 1 - \frac{1}{2} R_L^2 \left( 1 - \mu^2 \right) k_{\perp}^2 \right].
\label{2Dsmall1}
\edm
The first integral can be replaced by the diffusion coefficient of random walking magnetic field lines (see, e.g., Shalchi 2014)
\be
\kappa_{FL}^2 = \frac{\pi}{B_0^2} \int_{0}^{\infty} d k_{\perp} \; g^{2D}(k_{\perp}) k_{\perp}^{-2}
\ee
and the second one by the normalization condition (see, e.g., Shalchi 2009)
\be
\int_{0}^{\infty} d k_{\perp} \; g^{2D} (k_{\perp}) = \frac{\delta B^2}{2 \pi}.
\label{norm}
\ee
Therewith, Eq. (\ref{2Dsmall1}) becomes
\be
D_{\perp}^2 \approx \left( v \mu \right)^2 \kappa_{FL}^2 - \frac{1}{4} \left( v \mu \right)^2 \left( 1 - \mu^2 \right) R_L^2 \frac{\delta B^2}{B_0^2}.
\label{2dsmallcorr}
\ee
We can easily take the square root of the latter equation. By assuming that the second term in Eq. (\ref{2dsmallcorr}) is much smaller
than the first one, we can approximate the Fokker Planck coefficient of perpendicular diffusion by 
\be
D_{\perp} \approx v \left| \mu \right| \kappa_{FL} \left[ 1 - \frac{1}{8} \left( 1 - \mu^2 \right) \frac{R_L^2}{\kappa_{FL}^2} \frac{\delta B^2}{B_0^2} \right].
\label{DperpFLRWlimit}
\ee
The first term corresponds to the well-known field line random walk limit. In order to compute the perpendicular diffusion coefficient $\kappa_{\perp}$,
we have to combine Eqs. (\ref{kappaandDperp}) and (\ref{DperpFLRWlimit}) to find
\be
\kappa_{\perp} \approx \frac{v}{2} \kappa_{FL} \left[ 1 - \frac{1}{16} \frac{R_L^2}{\kappa_{FL}^2} \frac{\delta B^2}{B_0^2} \right].
\label{useflrwcoeff}
\ee
For spectrum (\ref{2dspec}) the field line diffusion coefficient was calculated in Shalchi \& Weinhorst (2009) based on the
Matthaeus et al. (1995) theory. They found
\be
\kappa_{FL} = \sqrt{\frac{s-1}{2(q-1)}} l_{\perp} \frac{\delta B}{B_0}
\label{FLresults}
\ee
as long as $q>1$ is satisfied. By combining Eq. (\ref{FLresults}) with (\ref{useflrwcoeff}), we can easily compute the
perpendicular diffusion coefficient or the perpendicular mean free path $\lambda_{\perp}=3 \kappa_{\perp}/v$. We derive
\be
\lambda_{\perp} \approx \frac{3}{2} \kappa_{FL} \left( 1 - \frac{q-1}{8 (s-1)} \frac{R_L^2}{l_{\perp}^2} \right).
\label{pmfpsmallr}
\ee
In this particular case the parameter $"a^2"$ occurring in non-linear diffusion theory has the value
\be
a^2 \approx 1 - \frac{q-1}{8 (s-1)} \frac{R_L^2}{l_{\perp}^2}.
\ee
Again we found that finite gyroradius effects reduce the perpendicular diffusion coefficient and, thus, $a^2 \leq 1$.
Furthermore, the parameter $"a^2"$ depends on the scale $l_{\perp}$ and the two spectral indexes $s$ and $q$.
\subsection{Numerical results for two-dimensional turbulence}
\begin{figure}
\centering 
\includegraphics[width=0.50\textwidth]{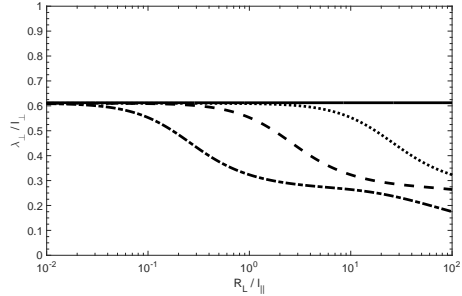}
\caption{The perpendicular mean free path $\lambda_{\perp}$ versus the unperturbed gyroradius $R_L$. The first parameter is normalized
with respect to the perpendicular scale $l_{\perp}$ and the second parameter to the parallel scale $l_{\parallel}$. All curves shown here
are obtained for non-linear transport in two-dimensional turbulence. We set the energy range spectral index $q=3$ to ensure that all
turbulence scales are finite. The solid line corresponds to the zero Larmor radius limit and the other curves are obtained by taking
into account the non-vanishing gyroradius. Shown are the results for $l_{\parallel} / l_{\perp} = 0.1$ (dotted line),
$l_{\parallel} / l_{\perp} = 1$ (dashed line), and $l_{\parallel} / l_{\perp} = 10$ (dash-dotted line).}
\label{2DNLT}
\end{figure}

For arbitrary Larmor radii $R_L$, we can compute the perpendicular mean free path only by solving Eq. (\ref{eq5p20}) numerically.
The results are visualized in Fig. \ref{2DNLT}. There, we have also shown the zero Larmor-radius result. The latter limit can be
obtained from Eq. (\ref{pmfpsmallr}) by setting $R_L=0$ therein. In this case the perpendicular mean free path becomes
\be
\frac{\lambda_{\perp}}{l_{\perp}} = \frac{3}{2} \frac{\kappa_{FL}}{l_{\perp}} = \frac{3}{2} \sqrt{\frac{s-1}{2(q-1)}} \frac{\delta B}{B_0}
\ee
where we have employed Eq. (\ref{FLresults}) also. For $s=5/3$, $q=3$, and $\delta B = B_0$ we find $\lambda_{\perp}/l_{\perp} \approx 0.612$.
Furthermore, we have computed $\lambda_{\perp}/l_{\perp}$ versus $R_L/l_{\parallel}$ for different values of the
ratio $l_{\parallel}/l_{\perp}$. In all considered cases we find that finite Larmor-radius corrections reduce the perpendicular
diffusion coefficient. For $l_{\perp} = 0.1 l_{\parallel}$, for instance, the perpendicular mean free path is about a factor $3$
smaller compared to the zero Larmor-radius limit. If the parameter $"a^2"$ is understood as finite gyroradius effect, this means that
$a^2 \approx 1/3$ for the case $l_{\perp} = 0.1 l_{\parallel}$. That is exactly in agreement with what was obtained by
Matthaeus et al. (2003).
\section{Summary and conclusion}
In the current paper we have investigated the effect of a finite gyroradius in the analytical theory of perpendicular diffusion.
To achieve a simplification of our calculations, we suppressed the non-linear effect of velocity diffusion. The case of strong velocity
diffusion will be explored in the second paper of the series.

In order to better understand finite gyroradius effects, we have combined two analytical theories for perpendicular diffusion
with different turbulence models. We considered quasi-linear and non-linear transport, respectively. For the turbulence we have
employed the slab model, a noisy slab model, and the two-dimensional model. The first model represents the case that the Kubo number
is zero, whereas the other two models represent turbulence with small and infinite Kubo numbers, respectively.

We have shown that finite gyroradius effects do not occur in slab turbulence as expected. For the other two models such effects can
be important depending on the size of the gyroradius. In all considered cases, finite gyroradius effects reduce the perpendicular
diffusion coefficient. A reduction of the perpendicular diffusion coefficient due to finite gyroradius effects was also found
in Neuer \& Spatschek (2006).

A reduction of the perpendicular diffusion coefficient can be important for describing particle acceleration at shock waves
(see, e.g., Ferrand et al. 2014). In such scenarios a small diffusion coefficient leads to a longer residence time of the particles
at the shock front and, thus, the particles experience acceleration to higher energies.

A fundamental problem in analytical treatments of perpendicular transport is the physical meaning and the value of the parameter
"a" as used in Eq. (\ref{introa}). One explanation of this parameter is provided in the current paper. The parameter discussed
here can be understood as finite gyroradius effect. In all cases found in the current paper, the finite gyroradius reduces the
perpendicular diffusion coefficient corresponding to a value $a^2 < 1$. This is basically what one finds if analytical results
are compared with test-particle simulations (see, e.g., Matthaeus et al. 2003 and Tautz \& Shalchi 2011). We have also shown that
the absolute value of $"a"$ depends  on the turbulence length scales $l_{\parallel}$ and $l_{\perp}$ as well as on the inertial
range spectral index $s$ and the energy range spectral index $q$. For two-dimensional turbulence and $l_{\perp} = 0.1 l_{\parallel}$,
for instance, we found $a^2 \approx 1/3$. This is exactly what was found before in Matthaeus et al. (2003). In the sequel of the current
paper we shall explore the case of strong velocity diffusion.
\section*{Acknowledgements}
{\it Support by the Natural Sciences and Engineering Research Council (NSERC) of Canada is acknowledged.}
\section*{Appendix: The Series $H(x,y)$}
In the current paper the series
\be
H (x, y) = \sum_{n=-\infty}^{+\infty} \frac{J_{n}^2 (x)}{y^2 + n^2}
\label{theseries}
\ee
occurs in the quasi-linear treatment of the transport in noisy slab turbulence (see Eq. (\ref{noisyperp1}))
and in the non-linear treatment of the transport in pure two-dimensional turbulence (see Eq. (\ref{eq5p21})).
Eq. (\ref{theseries}) is a special type of a so-called {\it Kapteyn series} (see Kapteyn 1893 and Watson 1966).
This special type of series was already discussed in Shalchi \& Schlickeiser (2004) where the following
asymptotic limits have been derived
\bdm
H(x \gg 1, y \ll 1)	& \approx &		\frac{J_0^2 (x)}{y^2}
\approx \frac{1}{\pi x y^2}, \nonumber\\
H(x \ll 1, y \ll 1)	& \approx &		\frac{1}{y^2},	\nonumber\\
H(x \gg y, y \gg 1)	& \approx &		\frac{1}{x y},	\nonumber\\
H(x \ll y, y \gg 1)	& \approx &		\frac{1}{y^2};
\label{eq5p23}
\edm
A very detailed and useful investigation of series (\ref{theseries}) can be found in Tautz \& Lerche (2010). These authors have shown
that the series can be written as
\be
H(x,y) = \frac{\pi J_{+iy} (x) J_{-iy} (x)}{y \sinh ( \pi y )}.
\label{TautzandLerche}
\ee
In the following we consider small and large arguments $x$ of the Bessel functions and simplify Eq. (\ref{TautzandLerche}) in these cases.
\subsection*{The limit $\left| x \right| \rightarrow 0$}
In the limit considered here, the Bessel functions can be approximated by a Taylor expansion (see, e.g., Watson 1966)
\be
J_{\nu} \left( x \right) = \sum_{n=0}^{\infty} \frac{\left( -1 \right)^{n}}{n! \Gamma \left( \nu + n + 1 \right)} \left( \frac{x}{2} \right)^{2n+\nu}
\ee
where we have used the Gamma function $\Gamma (z)$. The latter expansion is convergent for $|x| < \infty$ and arbitrary $\nu$.
Therewith, we can show that for small $x$
\bdm
J_{+iy} (x) J_{-iy} (x) & \approx & \frac{1}{\Gamma (1 + iy) \Gamma (1 - iy)} \nonumber\\
& \times & \left( 1 - \frac{1}{2} \frac{x^2}{1 + y^2} \right)
\label{besapro}
\edm
where we have also used the relation $\Gamma (z+1) = z \Gamma (z)$. Furthermore, we can employ (see, e.g., Abramowitz \& Stegun 1974)
\bdm
\Gamma (1 + iy) \Gamma (1 - iy) & = & i y \Gamma (iy) \Gamma (1 - iy) \nonumber\\
& = & \frac{\pi y}{-i \sin (i \pi y)} \nonumber\\
& = & \frac{\pi y}{\sinh ( \pi y )}.
\edm
With that result and by employing Eq. (\ref{besapro}), we can write Eq. (\ref{TautzandLerche}) as
\be
H(x,y) \approx \frac{1}{y^2} \left( 1 - \frac{1}{2} \frac{x^2}{1 + y^2} \right).
\label{Tautzlimit}
\ee
If we additionally assume that $y \ll 1$, we can approximate
\be
H(x,y) \approx \frac{1}{y^2} \left( 1 - \frac{1}{2} x^2 \right).
\label{Tautzlimit2}
\ee
If the term which is quadratic in $x$ is neglected, we obtain one of the limits listed in Eq. (\ref{eq5p23}). If $y \gg 1$,
Eq. (\ref{Tautzlimit}) would become
\be
H(x,y) \approx \frac{1}{y^2} \left( 1 - \frac{1}{2} \frac{x^2}{y^2} \right)
\label{Tautzlimit3}
\ee
and in lowest order $x/y$ we would still find the same limit as in Eq. (\ref{eq5p23}). 
\subsection*{The limit $\left| x \right| \rightarrow \infty$}
If the argument $x$ in the Bessel function is large, we can approximate (see, e.g., Watson 1966)
\be
J_{\nu} \left( x \right) \approx \sqrt{\frac{2}{\pi x}} \cos \left( x - \frac{\pi}{2} \nu - \frac{\pi}{4} \right).
\ee
Therefore, we find
\bdm
& & J_{+iy} \left( x \right) J_{-iy} \left( x \right) \nonumber\\
& \approx & \frac{2}{\pi x} \cos \left( x - \frac{\pi}{2} i y - \frac{\pi}{4} \right) \nonumber\\
& \times & \cos \left( x + \frac{\pi}{2} i y - \frac{\pi}{4} \right).
\edm
By using the relation
\be
\cos \left( a \pm b \right) = \cos \left( a \right) \cos \left( b \right) \mp \sin \left( a \right) \sin \left( b \right)
\ee
this becomes
\bdm
& & J_{+iy} \left( x \right) J_{-iy} \left( x \right) \nonumber\\
& \approx & \frac{2}{\pi x} \left[ \cos^2 \left( x - \frac{\pi}{4} \right) \cos^2 \left( \frac{\pi}{2} i y \right) \right. \nonumber\\
& - & \left. \sin^2 \left( x - \frac{\pi}{4} \right) \sin^2 \left( \frac{\pi}{2} i y \right) \right]
\edm
which can be written as
\bdm
& & J_{+iy} \left( x \right) J_{-iy} \left( x \right) \nonumber\\
& \approx & \frac{2}{\pi x} \left[ \cos^2 \left( \frac{\pi}{2} i y \right) - \sin^2 \left( x - \frac{\pi}{4} \right) \right].
\edm
The cosine function can be replaced by using $\cos (i z) = \cosh (z)$. Combining our findings with Eq. (\ref{TautzandLerche}), we obtain
\bdm
& & H (x,y) \approx \frac{2}{x y \sinh ( \pi y )} \nonumber\\
& \times & \left[ \cosh^2 \left( \frac{\pi}{2} y \right) - \sin^2 \left( x - \frac{\pi}{4} \right) \right].
\label{large1}
\edm
The latter approximation can be used in the limit $\left| x \right| \rightarrow \infty$. If we additionally assume that
$|y| \rightarrow 0$, Eq. (\ref{large1}) becomes
\be
H (x,y) \approx \frac{2}{\pi x y^2} \cos^2 \left( x - \frac{\pi}{4} \right)
\ee
in agreement with Eq. (\ref{eq5p23}). For $|y| \rightarrow \infty$, however, Eq. (\ref{large1}) becomes
\be
H (x,y) \approx \frac{1}{x y}
\ee
which is also listed in Eq. (\ref{eq5p23}).
{}

\end{document}